\documentclass[12pt]{article}

\usepackage[dvips]{graphicx}

\def\dspace{\baselineskip = 0.30in}

\def\lapproxeq{\lower .7ex\hbox{$\;\stackrel{\textstyle
<}{\sim}\;$}}
\def\gapproxeq{\lower .7ex\hbox{$\;\stackrel{\textstyle
>}{\sim}\;$}}

\begin{document}

\dspace

\begin{titlepage}
\begin{flushright}
BA-04-11\\
KIAS-P04039
\end{flushright}
\vskip 2cm
\begin{center}
{\Large\bf Inflation With Realistic Supersymmetric SO(10) } \vskip
1cm {\normalsize\bf $^{(a)}$Bumseok
Kyae\footnote{bkyae@kias.re.kr} and $^{(b)}$Qaisar
Shafi\footnote{shafi@bartol.udel.edu} } \vskip 0.5cm {\it
$^{(a)}$School of Physics, Korea Institute for Advanced Study,
\\207-43, Cheongnyangni-Dong, Dongdaemun-Gu, Seoul 130-722, Korea
\\$^{(b)}$Bartol Research Institute, University of
Delaware,
\\Newark,
DE~~19716,~~USA\\[0.1truecm]}

%

\end{center}
\vskip 2cm


\begin{abstract}

We implement inflation within a realistic supersymmetric $SO(10)$
model in which the doublet-triplet splitting is realized through
the Dimopoulos-Wilczek mechanism, the MSSM $\mu$ problem is
resolved, and higgsino mediated dimension five nucleon decay is
heavily suppressed. The cosmologically unwanted topological
defects are inflated away, and from $\delta T/ T$, the $B-L$
breaking scale is estimated to be of order $10^{16}-10^{17}$ GeV.
Including supergravity corrections, the scalar spectral index $n_s
= 0.99\pm 0.01$, with $|dn_s/ d{\rm ln}k| \lapproxeq 10^{-3}$.

\end{abstract}
\end{titlepage}

\newpage


In an attractive class of supersymmetric (SUSY) models inflation
is associated with spontaneous breaking of a gauge symmetry, such
that $\delta T/ T$ is proportional to $(M/M_{\rm Planck})^2$,
where $M$ denotes the symmetry breaking scale and $M_{\rm Planck}$
($\equiv 1.2\times 10^{19}$ GeV) denotes the Planck
mass~\cite{hybrid,review}. Thus, from measurements of $\delta
T/T$, $M$ is estimated to be of order $10^{16}$
GeV~\cite{hybrid,ns,dec}. The scalar spectral index $n_s$ in these
models is very close to unity\footnote{Following Ref.~\cite{dec},
that including supergravity corrections, these model can yield a
spectral index somewhat larger than unity.  It is in our current
model that $n_s=0.99\pm 0.01$, as we emphasize in the abstract. }
in excellent agreement with recent fits to the
data~\cite{wmap}.\footnote{Earlier there was some (weak) evidence
for a running spectral index with $dn_s/d{\rm ln}k\approx-5\times
10^{-2}$. But this is not confirmed by a more recent
analysis~\cite{recent}. We will consider the simplest models which
yield $dn_s/d{\rm ln}k\lapproxeq -10^{-3}$. However, more
complicated scenarios with two or more inflationary epoch can
yield a significantly larger $dn_s/d{\rm ln}k$~\cite{senoguz}.} A
$U(1)$ $R$-symmetry plays an essential role in the construction of
these inflationary models. These models possess another important
property, namely with the minimal K${\rm\ddot{a}}$hler potential,
the supergravity (SUGRA) corrections do not spoil the inflationary
scenario~\cite{review,sugrainf}, which has been realized with a
variety of attractive gauge groups including $SU(3)_c\times
SU(2)_L\times SU(2)_R\times U(1)_{B-L}$ ($\equiv
G_{LR}$)~\cite{LR}, $SU(4)_c\times SU(2)_L\times SU(2)_R$ ($\equiv
G_{422}$)~\cite{422} and $SU(5)$~\cite{su5}. [The gauge symmetries
$G_{LR}$ and $G_{422}$ were first introduced in
Refs.~\cite{lr,ps}.] Our goal in this paper is to implement
inflation in a realistic $SO(10)$ model.

$SO(10)$~\cite{so10} has two attractive features
that it shares with $G_{422}$, namely, it predicts the existence of right
handed neutrinos as well as the seesaw mechanism. These two features are
very helpful in understanding neutrino oscillations~\cite{kamio}
and also in generating a baryon asymmetry via leptogenesis~\cite{yanagida}.
Furthermore, at least within a four
dimensional setting, it seems easier to realize doublet-triplet (DT) splitting
without fine tuning in $SO(10)$
(say via the Dimopoulos-Wilczek mechanism~\cite{dw}) than in $SU(5)$.

To implement $SO(10)$ inflation we would like to work with a
realistic model with the following properties: DT splitting is
realized without fine tuning, and the low energy theory coincides
with the minimal supersymmetric standard model (MSSM). [For
$SO(10)$ inflation in a five dimensional setting, see
Ref.~\cite{5dinf}.] The MSSM $\mu$ problem should also be
resolved, and higgsino mediated dimension five nucleon decay
should be adequately suppressed.
%
Gauge boson mediated nucleon decay is still present with a
predicted nucleon lifetime of order $10^{34}-10^{36}$ yrs.
Finally, matter parity is unbroken, so that the LSP is stable and
makes up the dark matter in the universe.

To achieve natural DT splitting and the MSSM at low energies with
$SO(10)$, one is led to consider a non-minimal set of Higgs
superfields. This is to be contrasted with the subgroups of
$SO(10)$, such as $G_{LR}$ or $G_{422}$ above, in which the DT
splitting problem is absent.
%
%
Many authors have previously addressed the DT splitting and the
dimension five nucleon decay problem in
$SO(10)$~\cite{so10mssm,miniso10,dim5}, and the proposed solutions
are not necessarily straightforward. In this paper we will follow
Refs.~\cite{miniso10,dim5}, with suitable modifications needed to
make the scheme consistent with the desired inflationary scenario,
and also to avoid potential cosmological problems (monopoles,
moduli, etc). While doing this we would like to also ensure that
the SUGRA corrections also do not disrupt the inflationary
scenario.

%
%


A minimal set of Higgs required to break $SO(10)$ to the MSSM
gauge group $SU(3)_c\times SU(2)_L\times U(1)_Y$ ($\equiv G_{SM}$)
is ${\bf 45}_H$, ${\bf 16}_H$, ${\bf\overline{16}}_H$.  A non-zero
vacuum expectation value (VEV) of ${\bf 45}_H$ along the $B-L$
($I_{3R}$) direction breaks $SO(10)$ to $G_{LR}$ ($SU(4)_c\times
SU(2)_L\times U(1)_R$) and produces magnetic monopoles. The ${\bf
16}_H$, ${\bf\overline{16}}_H$ VEVs break $SO(10)$ to $SU(5)$ and
induce masses for the right handed neutrinos via dimension five
operators.
[Note that breaking of $SO(10)$ with ${\bf
16}_H+{\bf\overline{16}}_H$, in contrast to ${\bf 126}_H+{\bf
\overline{126}}_H$, does not produce $Z_2$ cosmic
strings~\cite{Z4}.]
One of our goals, of course, is to make sure that the topological
defects do not pose cosmological difficulties.  Thus, it would be
helpful if during inflation $SO(10)$ is, for instance, broken to
$G_{LR}$~\cite{earlier}, $SU(4)_c\times SU(2)_L\times U(1)_R$, or
$G_{SM}$.

To implement DT splitting without fine tuning and eliminate
dimension five proton decay, and to recover the MSSM at low
energies with the $\mu$ problem resolved, we need an additional
${\bf 45}$-plet (${\bf 45}_H'$), two additional ${\bf 16}+{\bf
\overline{16}}$ pairs, two ${\bf 10}$-plets (${\bf 10}_h$ and
${\bf 10}$), and several singlets~\cite{miniso10,dim5}. One more
${\bf 45}$-plet is also required by $U(1)$ $R$-symmetry. This
symmetry, among other things, plays an essential role in realizing
inflation, and its $Z_2$ subgroup coincides with the MSSM matter
parity. The $SO(10)$ singlet superfields are denoted as $S$, $X$,
$X'$, $Y$, $P$, $\overline{P}$, $Q$, and $\overline{Q}$, whose
roles will be described below. Table I displays the quantum
numbers (under the global $U(1)$ $R$ and $U(1)_A$ symmetries) of
all the Higgs sector superfields and the third family matter field
(${\bf 16}_3$). Following standard practice, we employ the same
notation for the superfields and their scalar components. \vskip
0.6cm
\begin{center}
\begin{tabular}{|c||cccccccccc|} \hline
 & $S$~ & $X$ & $X'$ & $Y$ & $P$ & $\overline{P}$ & $Q$ & $\overline{Q}$ &
  ${\bf 10}$ & ${\bf 10}_h$
 \\ \hline
$R$ & $1$ & $-1$ & $-1$ & $0$ & $0$ & $0$ & $-2$ & $2$ &
$1$ & $0$  \\
$A$ & $0$ & $-2/3$ & $-2/3$ & $-1/3$ & $-1/4$ & $1/4$ & $-1/2$ &
$1/2$ & $1/6$ & $0$
\\ \hline\hline
 & ${\bf 16}$ & ${\bf\overline{16}}$ & ${\bf 16}'$ & ${\bf \overline{16}}'$ &
${\bf 16}_H$ & ${\bf\overline{16}}_H$ & ~${\bf 16}_3$~ & ${\bf
45}$ & ${\bf 45}_H$ & ${\bf 45}_H'$
\\ \hline
$R$ & $1$ & $3$ & $2$ & $2$ & $0$ & $0$ & $1/2$ & $1$ & $0$ & $-1$

\\
$A$ & $1/2$ & $~2/3~$ & $2/3$ & $2/3$ & $0$ & $0$ & ~$0$~ & $1/2$
& $-1/6$ & $-1/3$
\\ \hline
\end{tabular}
\vskip 0.4cm
{\bf Table I~}
\end{center}

To break $SO(10)$ to $G_{LR}$, consider the superpotential,
\begin{eqnarray} \label{adj}
&&~~~~~~~~~~~~ W_{45}= \frac{\alpha}{6M_*} X^{(')}Y{\rm
Tr}\bigg({\bf 45}{\bf 45}\bigg) -\frac{\beta}{6} Y{\rm
Tr}\bigg({\bf 45}{\bf 45}_H\bigg)
\\
&&~ +\frac{\gamma_1}{36M_*}{\rm Tr}\bigg({\bf 45}{\bf 45}_H\bigg)
{\rm Tr}\bigg({\bf 45}_H{\bf 45}_H\bigg)
+\frac{\gamma_2}{6M_*}{\rm Tr}\bigg({\bf 45}{\bf 45}_H{\bf
45}_H{\bf 45}_H \bigg)
\nonumber ~,
\end{eqnarray}
where $\alpha$, $\beta$, $\gamma_{1,2}$ are dimensionless
parameters, and $M_*$ $(\sim 10^{18}$ GeV) denotes the cutoff
scale. As will be explained, $X$, $X'$, and $Y$ can develop
non-zero VEVs, $\langle X\rangle \sim \langle X'\rangle \sim
\langle Y\rangle \sim 10^{16}$ GeV. Due to non-zero $\langle
Y\rangle$, ${\bf 45}_H$ can also obtain a VEV in the $B-L$
direction from the $\beta$ and $\gamma_{1,2}$ terms of
Eq.~(\ref{adj}),
\begin{eqnarray} \label{vev2}
\langle {\bf 45}_{H}\rangle =\left(\begin{array}{ccc|cc}
v&&&& \\
&v&&& \\
&&v&& \\ \hline
&&&0& \\
&&&&0
\end{array}\right)\otimes i\sigma_2 ~,~~~{\rm and}~~~
\langle {\bf 45}\rangle =0 ~,
\end{eqnarray}
where $v=\sqrt{\frac{\beta}{\gamma}\langle Y\rangle M_*}\equiv
M_{GUT}$ ($\approx 3\times 10^{16}$ GeV), with $\gamma\equiv
\gamma_1+\gamma_2$. The $3\times 3$ block corresponds to $SU(3)_c$
and the $2\times 2$ block to $SU(2)_L$ of the MSSM gauge group.
Hence, the $SO(10)$ gauge symmetry is broken to $G_{LR}$. Note
that from the `$\alpha$ term,' the ${\bf 45}$ multiplet becomes
superheavy. It acquires a VEV of order $(m_{3/2}M_{GUT})/M_*$
after SUSY breaking, where $m_{3/2}$ ($\sim$ TeV) denotes the
scale of the soft parameters.

The next step in the breaking
to the MSSM gauge group $G_{SM}$ ($=G_{LR}\cap SU(5)$)
is achieved with the following superpotential,
\begin{eqnarray} \label{spinor}
&& ~ W_{16} =S\bigg[\kappa {\bf 16}_H{\bf\overline{16}}_H+\lambda
{\bf 10}_h{\bf 10}_h - \kappa M_{B-L}^2\bigg]
-\frac{\rho}{M_*^2}S({\bf 16}_H{\bf\overline{16}}_H)^2  ~
\nonumber \\
&& +{\bf 16}\bigg[\frac{\lambda_1}{M_*}{\bf
45}_HY-\frac{\lambda_2}{M_*}P^2 \bigg]{\bf\overline{16}}_H
+{\bf\overline{16}}\bigg[\frac{\lambda_3}{M_*}{\bf
45}_HQ-\frac{\lambda_4}{M_*}
({\bf 45}_H^{'})^2\bigg]{\bf 16}_H  \\
&&
~~
 + {\bf 16}'\bigg[\frac{\lambda_5}{M_*}{\bf
45}_H'Y-\lambda_6X
\bigg]{\bf\overline{16}}_H+{\bf\overline{16}}'\bigg[\frac{\lambda_7}{M_*}{\bf
45}_H'Y-\lambda_8X'\bigg]{\bf 16}_H ~, \nonumber
\end{eqnarray}
where $\rho$ is a dimensionless coupling constant. The
dimensionful parameter $M_{B-L}$, as determined from inflation
($\delta T/T$), turns out to be of order $10^{16}-10^{17}$
GeV~\cite{dec}. The superfield ${\bf 10}_h$ includes the two MSSM
Higgs doublets. As previously mentioned, additional ${\bf 16}$,
${\bf\overline{16}}$ are essential to stabilize the VEV of ${\bf
45}_H$ in Eq.~(\ref{vev2})~\cite{miniso10}. From the $\kappa$ and
$\rho$ terms, ${\bf 16}_H$ and ${\bf \overline{16}}_H$ develop
VEVs
%
of order $M_{B-L}$, breaking $SO(10)$ to $SU(5)$,
\begin{eqnarray} \label{vev3}
|\langle{\bf 16}_H\rangle|^2
=|\langle{\bf\overline{16}}_H\rangle|^2=
\frac{M_{B-L}^2}{2\zeta}\bigg[1-\sqrt{1-4\zeta}\bigg] ~,
\end{eqnarray}
where $\zeta\equiv\rho M_{B-L}^2/(\kappa M_*^2)$~\cite{422}, while
$\langle S\rangle =\langle{\bf 10}_h\rangle=0$ upto corrections of
$O(m_{3/2})$ by including soft SUSY breaking terms in the scalar
potential~\cite{LR}. The ``D-term'' scalar potential vanishes
along the (D-flat) direction $|\langle{\bf 16}_H\rangle|
=|\langle{\bf\overline{16}}_H^{~*}\rangle|$
($=|\langle{\bf\overline{16}}_H\rangle|$). Together with
Eq.~(\ref{vev2}), the $SO(10)$ gauge symmetry is broken to the
MSSM gauge symmetry.
The MSSM Higgs doublets arise from ${\bf 10}_h$. With $\langle
S\rangle\approx -m_{3/2}/\kappa$, the $\mu$ term from
Eq.~(\ref{spinor}) is of order $(\lambda/\kappa)m_{3/2}\sim$ TeV,
for $\kappa\approx \lambda$.\footnote{ From the non-renormalizable
term $y_{\mu}{\bf 10}{\bf 10}_h\langle {\bf 16}_H{\bf 45}_H{\bf
\overline{16}}_H\rangle/M_*^2$, the doublets in ${\bf 10}_h$
obtains a ``seesaw mass'' $y_\mu^2(\langle {\bf 16}_H{\bf
45}_H{\bf \overline{16}}_H\rangle)^2/(M_*^4\langle {\bf
45}_H'\rangle)\sim$ TeV with $y_\mu \sim 10^{-3}$, which modifies
the $\mu$ parameter at low energies.} Similarly the soft term
$B\mu$ ($\approx -2(\lambda/\kappa)m_{3/2}^2$) is
generated~\cite{LR,king}.

Our next step is to ensure that the low energy theory coincides
precisely with the MSSM. With $SO(10)$ broken to $G_{LR}$ by
$\langle{\bf 45}_H\rangle$ via Eq.~(\ref{adj}), the goldstone
modes from ${\bf 45}_H$, $[\{{\bf (3,\overline{2})}_{-5/6}$, ${\bf
(3,2)}_{1/6}$, ${\bf (\overline{3},1)}_{-2/3}\}+{\rm h.c.}]$ in
terms of $G_{SM}$, are absorbed by the gauge sector. The states of
${\bf (8,1)}_0$, ${\bf (1,3)}_0$, ${\bf (1,1)}_0$, ${\bf
(1,1)}_{1}$, and ${\bf (1,1)}_{-1}$ contained in ${\bf 45}_H$
acquire superheavy masses through the quartic couplings. On the
other hand, when $SO(10)$ breaks to $SU(5)$ by $\langle {\bf
16}_H\rangle$ and $\langle {\bf\overline{16}}_H\rangle$, the
states $[\{{\bf (3,2)}_{1/6}$, ${\bf (\overline{3},1)}_{-2/3}$,
${\bf (1,1)}_{1}\}+{\rm h.c.}]+{\bf (1,1)}_{0}$ in ${\bf 16}_H$,
${\bf\overline{16}}_H$ should be absorbed by the gauge sector,
while $[\{{\bf (\overline{3},1)}_{1/3}$, ${\bf
(1,\overline{2})}_{-1/2}\}+{\rm h.c.}]$ remain massless (or
light). Note that $[\{{\bf (3,2)}_{1/6}$, ${\bf
(\overline{3},1)}_{-2/3}\}+{\rm h.c.}]$ are common between ${\bf
45}_H$ and ${\bf 16}_H$, ${\bf\overline{16}}_H$. Thus, when
$SO(10)$ breaks to $G_{SM}$ by an adjoint and a vector-like pair
of spinorial Higgs, the superfields associated with $[\{{\bf
(3,2)}_{1/6}$, ${\bf (\overline{3},1)}_{-2/3}$, ${\bf
(\overline{3},1)}_{1/3}$, ${\bf (1,\overline{2})}_{-1/2}\}+{\rm
h.c.}]$ are pseudo-goldstone modes.
%
%
The extra light multiplets would spoil the unification of the MSSM
gauge couplings, and therefore must be eliminated.

The simplest way to remove them from the low energy spectrum is to
introduce couplings such as ${\bf 16}_H{\bf
45}_H{\bf\overline{16}}_H$. However, the presence of such a term
in the superpotential destabilizes the form of $\langle {\bf
45}_H\rangle$ given in Eq.~(\ref{vev2}), in such a way that at the
SUSY minimum, $v=0$ is required. It was shown in
Ref.~\cite{miniso10} that with the `$\lambda_i$' couplings
($i=1,2,3,4$) and an additional ${\bf 16}$--${\bf\overline{16}}$
pair in Eq.~(\ref{spinor}), the unwanted pseudo-goldstone modes
all become superheavy, keeping intact the form of Eq.~(\ref{vev2})
at the SUSY minimum.

From the ``F-flat conditions'' with ${\bf 16}_H$ and
${\bf\overline{16}}_H$ acquiring non-zero VEVs, one finds
\begin{eqnarray} \label{vev4}
&& \langle {\bf 45}_H\rangle\langle Y\rangle
=\frac{\lambda_2}{\lambda_1}\langle P^2\rangle ~,~~~ \langle {\bf
45}_H\rangle\langle Q\rangle =\frac{\lambda_4}{\lambda_3}{\rm
Tr}\langle {\bf 45}_H'\rangle^2 ~.
\end{eqnarray}
Thus, if $P$ and $Q$
develop VEVs, $\langle{\bf 45}_H\rangle$, $\langle {\bf
45}_H'\rangle$, and $\langle Y\rangle$ should also appear. We will
soon explain how $\langle P\rangle$ and $\langle Q\rangle$ arise.
Since $\langle Y\rangle$ is related to $\langle {\bf 45}_H\rangle$
via Eq.~(\ref{vev2}), both are uniquely determined. We assume that
$\langle {\bf 45}_H'\rangle$ points in the $I_{3R}$ direction,
\begin{eqnarray}
\langle {\bf 45}_{H}'\rangle =\left(\begin{array}{ccc|cc}
0&&&& \\
&0&&& \\
&&0&& \\ \hline
&&&v'& \\
&&&&v'
\end{array}\right)\otimes i\sigma_2 ~.
\end{eqnarray}
Recall that $\langle {\bf 45}_H'\rangle$ is employed to suppress
higgsino mediated dimension five nucleon decay~\cite{dim5}.
Similarly, due to the presence of the `$\lambda_i$' ($i=5,6,7,8$)
couplings in Eq.(\ref{spinor}), the low energy spectrum is
protected even with the ${\bf 45}_H'$ present~\cite{dim5}. With
non-zero VEVs for ${\bf 45}_H'$ and $Y$, $X$ and $X'$ slide to
values satisfying
\begin{eqnarray} \label{vev5}
\frac{\lambda_{5,7}}{M_*}\langle {\bf 45}_H'\rangle\langle
Y\rangle-\lambda_{6,8}\langle X^{(')}\rangle=0 ~,
\end{eqnarray}
with $|\langle {\bf 16}'\rangle|=|\langle {\bf
\overline{16}}\rangle|\sim O(m_{3/2})$.
%
In order to guarantee the `$\lambda_i$' couplings in
Eq.~(\ref{spinor}) and to forbid ${\bf 16}_H{\bf
45}_H{\bf\overline{16}}_H$,  the $U(1)$ symmetries in Table I are
essential.

To obtain non-vanishing VEVs for $P$ and $Q$, one could, as a
simple example, consider the following superpotential,
\begin{eqnarray} \label{PQ}
W_{PQ}= S\bigg[\kappa_1P\overline{P}+\kappa_2Q\overline{Q}\bigg]
-\frac{S}{M_*^2}\bigg[\rho_1(P\overline{P})^2+\rho_2(Q\overline{Q})^2\bigg]~,
\end{eqnarray}
such that
\begin{eqnarray} \label{pqv}
\langle P\overline{P}\rangle=\frac{\kappa_1}{\rho_1}M_*^2\sim
M_{GUT}^2~,~~~\langle
Q\overline{Q}\rangle=\frac{\kappa_2}{\rho_2}M_*^2\sim M_{GUT}^2 ~.
\end{eqnarray}
The $\lambda_{2,3}$ terms in Eq.~(\ref{spinor}) just determine
$\langle {\bf 45}_H\rangle$, $\langle Y\rangle$, and $\langle{\bf
45}_H'\rangle$. With the inclusion of soft SUSY breaking terms,
the VEVs $\langle P\rangle$, $\langle\overline{P}\rangle$,
$\langle Q\rangle$, and $\langle\overline{Q}\rangle$ would be
completely fixed. To avoid potential cosmological problems
associated with moduli fields, we make the important assumption
that the VEVs satisfy the constraints $\langle
P\rangle=\langle\overline{P}\rangle$ and $\langle
Q\rangle=\langle\overline{Q}\rangle$. This could be made plausible
by assuming universal soft scalar masses, and that the SUSY
breaking ``A-terms" asymmetric under
$P\leftrightarrow\overline{P}$ and $Q\leftrightarrow\overline{Q}$
are plausibly small enough.\footnote{In gravity mediated SUSY
breaking scenario with the minimal K${\rm\ddot{a}}$hler potential,
``A-terms'' are given by
$m_{3/2}\times[(A-3)W+\sum_i\phi_i\frac{\partial W}{\partial
\phi_i}+{\rm h.c.}]$, where $A$ is a dimensionless number
associated with hidden sector dynamics~\cite{prt}.  Since
dimensions of the operators associated with the $\lambda_k$'s
($k=1,2,3,4$) in Eq.~(\ref{spinor}) are all the same, the
``A-term'' coefficients ($\equiv A_{\lambda_k}$) corresponding to
$\lambda_k$ should be $m_{3/2}(A+1)\lambda_k$, and so satisfy
$A_{\lambda_{j+1}}/A_{\lambda_{j}}=\lambda_{j+1}/\lambda_{j}$
($j=1,3$). Hence, at the minimum, the ``A-terms'' corresponding to
$\lambda_{k}$ are cancelled by each other with the VEVs in
Eq.~(\ref{vev4}). Since the other soft terms are symmetric under
$P\leftrightarrow\overline{P}$ and $Q\leftrightarrow\overline{Q}$,
we have $\langle P\rangle=\langle\overline{P}\rangle$ and $\langle
Q\rangle=\langle\overline{Q}\rangle$ at the minimum of the scalar
potential. }$^{,}$\footnote{In gauge mediated SUSY breaking
scenario, ``A-terms'' are generally suppressed.} Note that even
with the soft SUSY breaking terms in the Lagrangian, the GUT scale
results Eqs.~(\ref{vev4}) and (\ref{pqv}) should be still
effectively valid. Since the fields that couple to $P$,
$\overline{P}$, $Q$ and $\overline{Q}$ are all superheavy, the
soft parameters are expected to be radiatively stable at low
energies.
%
%
Thus, at the minimum of the scalar potential, we have four mass
eigen states, $(P\pm\overline{P})/\sqrt{2}$ ($\equiv P_{\pm}$) and
$(Q\pm\overline{Q})/\sqrt{2}$ ($\equiv Q_{\pm}$). While $P_+$ and
$Q_+$ obtain superheavy masses of order $M_{GUT}$ and large VEVs (
$=\sqrt{\kappa_{1,2}/\rho_{1,2}}M_*+O(m_{3/2})\sim M_{GUT}$,
respectively), $P_-$ and $Q_-$ remain light ($\sim m_{3/2}$) with
vanishing VEVs.
%

With $\langle{\bf 45}_H\rangle$ in Eq.~(\ref{vev2}), the ``DT
splitting problem'' resolves itself through the mechanism
in~\cite{dw}. Consider the superpotential:
\begin{eqnarray} \label{vec}
W_{10}=y_1{\bf 10}{\bf 45}_H'{\bf 10}+y_2{\bf 10}{\bf 45}_H{\bf
10}_h ~.
\end{eqnarray}
From the first term in Eq.~(\ref{vec}), only the doublets
contained in ${\bf 10}$ become superheavy~\cite{dim5}, and from
the second term only the color triplet fields included in ${\bf
10}$ and ${\bf 10}_h$ acquire superheavy
masses~\cite{dw,miniso10,dim5}.  Since the two color triplets
contained in ${\bf 10}_h$ do not couple in Eq.~(\ref{vec}),
dimension five nucleon decay which may be in conflict with the
Superkamiokande observations~\cite{kamio} is eliminated in the
SUSY limit~\cite{dim5}.  Note that operators such as ${\bf 10}{\bf
10}_h$, ${\bf 10}_h{\bf 10}_h$, $[{\bf 10}{\bf 10}_h]{\rm Tr}({\bf
45}_H{\bf 45}_H)$ and so on are allowed by $SO(10)$ and, unless
forbidden, would destroy the gauge hierarchy. The $U(1)$
symmetries in Table I are once again crucial in achieving this.

Although the superpotential coupling $\langle S\rangle{\bf
10}_h{\bf 10}_h$ induces higgsino mediated dimension five nucleon
decay, there is a huge suppression factor of $m_{3/2}/M_{GUT}$.
Thus, we expect that nucleon decay is dominated by the exchange of
the superheavy gauge bosons with an estimated lifetime
$\tau_p\rightarrow e^+ \pi^0$ of order $10^{34}-10^{36}$ yrs. Note
that we have assumed that dimension five operators such as ${\bf
16}_i{\bf 16}_j{\bf 16}_k{\bf 16}_l$, ${\bf 16}_i{\bf 16}_j{\bf
16}_k{\bf 16}_H$ and so on, where the subscripts are family
indices of the matter, are adequately suppressed by assigning
suitable $R$ and $A$ charges to these matter superfields. This is
closely tied to the flavor problem, which we will not address
here.

Consider next the superpotential couplings involving the third generation
matter superfields,
\begin{eqnarray} \label{matter}
W_m=y_3{\bf 16}_3{\bf 16}_3{\bf 10}_h +\frac{y_{\nu}}{M_*} {\bf
16}_3{\bf 16}_3{\bf\overline{16}}_H{\bf\overline{16}}_H ~.
\end{eqnarray}
The first term yields Yukawa unification so that the MSSM
parameter ${\rm tan}\beta\approx m_t/m_b$~\cite{tanB}. For a
realistic construction of the fermion's mass matrices in $SO(10)$,
refer to e.g. Ref.~\cite{abb}.
%
From the $y_\nu$ term, the right handed neutrino masses are
$\lapproxeq y_\nu M_{B-L}^2/M_*\sim 10^{14}$ GeV. Right handed
neutrino masses of order $10^{14}$ GeV and smaller can yield a
mass spectrum for the light neutrinos through the seesaw
mechanism, that is suitable for neutrino oscillations. These
masses are also appropriate for realizing leptogenesis after
inflation~\cite{lepto-inf,senoguz}. Finally let us note that the
${\bf 16}_H$, ${\bf\overline{16}}_H$ VEVs break the center $Z_4$
of $SO(10)$ completely~\cite{Z4}. The role of `matter parity' is
played by the unbroken $Z_2$ subgroup of the $U(1)$
$R$-symmetry~\cite{LR}.  Thus the LSP in our model is expected to
be stable and contribute to the dark matter in the universe.

For completeness, we need to present also the other possible terms
in the superpotential that were not discussed in Eqs.~(\ref{adj}),
(\ref{spinor}), (\ref{PQ}), (\ref{vec}), and (\ref{matter}).
Indeed, we have more quartic couplings; ${\bf 10}{\bf 45}{\bf
10}_hX$, ${\bf 16}{\bf \overline{16}}XQ$, ${\bf 16}{\bf 16}_H{\bf
10}X$, ${\bf 16}'{\bf 16}_H{\bf 10}_hX$, ${\bf 16}_H{\bf 16}_H{\bf
10}_hS$, ${\bf 16}_H{\bf 16}_H{\bf 10}{\bf 45}_H$, ${\bf
\overline{16}}'{\bf \overline{16}}_H{\bf 10}_hX$, ${\bf
\overline{16}}_H{\bf \overline{16}}_H{\bf 10}_hS$, ${\bf
\overline{16}}_H{\bf \overline{16}}_H{\bf 10}{\bf 45}_H$, and so
on, which also are consistent with the charge assignment in Table
I. But they just provide sub-dominant effects in this model. For
instance, ${\bf 16}{\bf \overline{16}}\langle XQ\rangle$ can not
change the symmetry breaking pattern discussed above, because
$\langle {\bf 16}\rangle =\langle {\bf \overline{16}}\rangle =0$.
Thus, with keeping massless goldstones, it just modifies the
masses of the pseudo-goldstone modes contained in ${\bf 16}$ and
${\bf \overline{16}}$. Due to $\langle {\bf 16}_H\rangle\neq 0$
and $\langle {\bf 45}_H\rangle\neq 0$, ${\bf 16}_H{\bf 16}_H{\bf
10}{\bf 45}_H$ slightly changes masses of the $SU(2)_L$ doublets
contained in ${\bf 16}_H$ and ${\bf 10}$.  We also have over forty
extra penta-couplings except those considered in
Eqs.~(\ref{spinor}) and (\ref{PQ}), but we will neglect them.


Let us now discuss how inflation is implemented in the model
described so far. In particular, we aim to show that the SUGRA
corrections do not significantly affect the inflationary scenario,
which is a non-trivial result in inflationary model building. The
``F-term'' scalar potential in SUGRA is given by
\begin{eqnarray} \label{scalarpot}
V_F=e^{K/M_P^2}\bigg[\sum_{i,j}(K^{-1})^i_j(D_{\phi_i}W)(D_{\phi_j}W)^{*}
-3\frac{|W|^2}{M_P^2}\bigg] ~,
\label{scalarpot2}
\end{eqnarray}
where $M_P$ ($\equiv M_{\rm Planck}/\sqrt{8\pi}=2.4\times 10^{18}$
GeV) denotes the reduced Planck mass. $K$
($=K(\phi_i,\phi^*_j)=K^*$) and $W$ ($=W(\phi_i)$) are the
K${\rm\ddot{a}}$hler potential and the superpotential,
respectively. $(K^{-1})^i_j$ in Eq.~(\ref{scalarpot}) denotes the
inverse of $\partial^2 K/\partial\phi_i\partial\phi^*_j$. In our
case, $W$ is composed of Eqs.~(\ref{adj}), (\ref{spinor}),
(\ref{PQ}), (\ref{vec}), and (\ref{matter}). $D_{\phi_i}W$ in
Eq.~(\ref{scalarpot}) is defined as
\begin{eqnarray}
D_{\phi_i}W\equiv\frac{\partial W}{\partial\phi_i} +\frac{\partial
K}{\partial\phi_i}\frac{W}{M_P^2} ~.
\end{eqnarray}
The K${\rm\ddot{a}}$hler potential can be expanded as
$K=|\phi_i|^2+c_4|\phi_i|^4/M_P^2+\cdots$. For simplicity, we
consider the minimal case with $\partial^2
K/\partial\phi_i\partial\phi^*_j=\delta^i_j$. Indeed, as explained
in~\cite{review}, higher order terms in $K$ (with a coefficient
$\lapproxeq 10^{-2}$ for the quartic term) do not seriously affect
inflation. For simplicity, we will also ignore the TeV scale
electroweak symmetry breaking effects when discussing inflation.

In this paper, we aim to employ the `shifted' hybrid inflationary
scenario proposed in Ref~\cite{422}, in which symmetries can be
broken during inflation unlike the simple ``hybrid inflation''
model~\cite{hybrid}. An inflationary scenario is realized in the
early universe with the scalar fields $S$, ${\bf 16}_H$, ${\bf
\overline{16}}_H$, $P$, $\overline{P}$, $Q$, and $\overline{Q}$
displaced from the present values. We suppose that initially
$|\langle S\rangle|^2 \gapproxeq M_{B-L}^2[1/(4\zeta)-1]/2$ with
$1/4<\zeta<1/7.2$~\cite{422}, and $\langle{\bf 16}_H\rangle$,
$\langle{\bf\overline{16}}_H\rangle$, $\langle P\rangle$,
$\langle\overline{P}\rangle$, $\langle Q\rangle$, $\langle
\overline{Q}\rangle \neq 0$ with the inflationary superpotential
given by~\cite{422},
\begin{eqnarray}
W_{\rm infl}&\approx& -\kappa S\bigg[ M_{B-L}^2- {\bf
16}_H{\bf\overline{16}}_H+\frac{\rho}{\kappa M_*^2 }({\bf
16}_H{\bf\overline{16}}_H)^2
\nonumber\\
&&-\frac{\kappa_1}{\kappa}P\overline{P}+\frac{\rho_1}{\kappa
M_*^2}(P\overline{P})^2
-\frac{\kappa_2}{\kappa}Q\overline{Q}+\frac{\rho_2}{\kappa M_*^2}
(Q\overline{Q})^2 \bigg]
\\
&\equiv& -\kappa SM_{\rm eff}^2 ~, \nonumber
\end{eqnarray}
where $M_{\rm eff}^2$ turns out to be of order $M_{B-L}^2$.  With
$D_SW\approx -\kappa M_{\rm eff}^2(1+|S|^2/M_P^2)$, one can see
that the ``F-term'' scalar potential becomes
\begin{eqnarray} \label{inflpot}
V_{F}\approx\bigg(1+\sum_{k}\frac{|\phi_k|^2}{M_P^2}+\cdots\bigg)\bigg[
\kappa^2M_{\rm eff}^4\bigg(1+\frac{|S|^4}{2M_P^4}\bigg)
+\bigg(1+\frac{|S|^2}{M_P^2}+\frac{|S|^4}{2M_P^4}\bigg)
\sum_{k}|D_{\phi_k}W|^2\bigg]
~,
\end{eqnarray}
where all scalar fields except $S$ contribute to $\phi_k$. The
factor $(1+\sum_{k}|\phi_k|^2/M_P^2+\cdots)$ in front originates
from $e^{K/M_P^2}$ in Eq.~(\ref{scalarpot}).
%
%
In Eq.~(\ref{inflpot}) the quadratic term in $S$ from $|D_SW|^2$,
which is of order $(\kappa^2M_{\rm eff}^4/M_P^2)|S|^2$ ($\approx
H^2|S|^2$), has canceled out with the factor ``$-3|W|^2/M_P^2$''
($\approx -3\kappa^2M_{\rm eff}^4|S|^2/M_P^2$) and the quadratic
term in $S$ from ``$e^{K/M_P^2}$'' ($=1+|S|^2/M_P^2+\cdots$). It
is a common feature in this class of models~\cite{review}. Thus,
only if $|D_{\phi_k}W|/M_P$'s are much smaller than the Hubble
scale ($\sim \kappa M_{\rm eff}^2/M_P $), the flatness of $S$ will
be guaranteed even with the SUGRA corrections included. Note that
the $U(1)$ $R$-symmetry ensures the absence of terms proportional
to $S^2$, $S^3$, etc. in the superpotential, which otherwise could
spoil the slow-roll conditions.


Let us consider the inflationary trajectory on which $\langle{\bf
10}\rangle =\langle{\bf 10}_h\rangle=\langle{\bf
16}\rangle=\langle{\bf \overline{16}}\rangle =\langle{\bf
16}'\rangle=\langle{\bf \overline{16}}'\rangle =\langle{\bf
16}_3\rangle= \langle{\bf 45}\rangle=0$, with $D_{\bf 10}W=D_{{\bf
10}_h}W=D_{\bf 16}W=D_{\bf\overline{16}}W =D_{\bf
16'}W=D_{\bf\overline{16}'}W=D_{{\bf 16}_3}W=D_{{\bf 45}}W=0$.  On
the other hand,
\begin{eqnarray}
D_{{\bf 16}_H}W&=&\kappa S\bigg[{\bf
\overline{16}}_H\bigg(1-\frac{2\rho}{\kappa M_*^2 }({\bf
16}_H{\bf\overline{16}}_H)^2\bigg) -{\bf 16}_H^*\frac{M_{\rm
eff}^2}{M_P^2}\bigg] ~,
\\
D_{P}W&=&\kappa S\bigg[
\overline{P}\bigg(\frac{\kappa_1}{\kappa}-\frac{2\rho_1}{\kappa
M_*^2 }(P\overline{P})^2\bigg) -P^*\frac{M_{\rm
eff}^2}{M_P^2}\bigg] ~,
\\
D_{Q}W&=&\kappa
S\bigg[\overline{Q}\bigg(\frac{\kappa_2}{\kappa}-\frac{2\rho_2}{\kappa
M_*^2 }(Q\overline{Q})^2\bigg) -Q^*\frac{M_{\rm
eff}^2}{M_P^2}\bigg] ~,
\end{eqnarray}
and similarly $D_{{\bf \overline{16}}_H}W=D_{{\bf 16}_H}W({\bf
16}_H\leftrightarrow {\bf \overline{16}}_H)$,
$D_{\overline{P}}W=D_PW(P\leftrightarrow \overline{P})$, and
$D_{\overline{Q}}W=D_QW(Q\leftrightarrow \overline{Q})$. The other
$D_{\phi_l}W$'s ($\phi_l=X^{(')}$, $Y$, ${\bf 45}_H$, ${\bf
45}_H'$) are approximately given by $-s \langle\phi_l^*\rangle$,
where $s\equiv -W/M_P^2\approx \kappa \langle S\rangle M_{\rm
eff}^2/M_P^2$ ($<<M_{GUT}$). At one of the local minima,
$\langle{\bf 16}_H\rangle$, $\langle{\bf \overline{16}}_H\rangle$,
$\langle P\rangle$, $\langle\overline{P}\rangle$, $\langle
Q\rangle$, $\langle\overline{Q}\rangle$, and the vacuum energy
$V_{0}^{1/4}$ acquire the following values,
\begin{eqnarray}\label{sugravev}
&& |\langle {\bf
16}_H\rangle|^2=|\langle{\bf\overline{16}}_H\rangle|^2 \approx
\frac{\kappa
M_*^2}{2\rho}\bigg[1-\frac{M_{B-L}^2}{M_P^2}+\frac{\kappa
M_*^2}{2\rho M_P^2
}\bigg(1-\frac{M_{B-L}^2}{4S^2}+O(\kappa_{1,2}^2/\kappa^2) \bigg)
\bigg] , ~~~~~
\\
\label{pvev} &&|\langle
P\rangle|^2=|\langle\overline{P}\rangle|^2\approx \frac{\kappa_1
M_*^2}{2\rho_1}\bigg[1-\frac{\kappa M_{B-L}^2}{\kappa_1M_P^2}
+\frac{\kappa^2M_*^2}{4\kappa_1\rho M_P^2}
\bigg(1+O(\kappa_{1,2}^2/\kappa^2)\bigg)
\bigg]~, ~~
\\ \label{qvev}
&& |\langle Q\rangle|^2=|\langle\overline{Q}\rangle|^2\approx
\frac{\kappa_2 M_*^2}{2\rho_2}\bigg[1-\frac{\kappa
M_{B-L}^2}{\kappa_2M_P^2}
+\frac{\kappa^2M_*^2}{4\kappa_2\rho
M_P^2} \bigg(1+O(\kappa_{1,2}^2/\kappa^2)\bigg) \bigg]~, ~~
\\ \label{vac}
&& V_{0}\approx \kappa^2M_{0}^4\bigg[1+
\frac{M_*^2}{M_P^2}\bigg(\frac{\kappa}{\rho}+\frac{\kappa_1}{\rho_1}
+\frac{\kappa_2}{\rho_2}+O(\kappa^2,\kappa M_{B-L}^2/M_P^2)\bigg)
+ \sum_l\frac{|\langle\phi_l\rangle|^2}{M_P^2}\bigg] ~ ,
\end{eqnarray}
where we assumed
$\frac{\kappa}{\rho}\gapproxeq\frac{\kappa_1}{\rho_1}$, $
\frac{\kappa_2}{\rho_2}$ with $\kappa<<1$ and
$\rho\sim\rho_1\sim\rho_2\sim O(1)$. In Eq.~(\ref{vac}),
$M_{0}^4\equiv
M_{B-L}^4[1/(4\zeta)+1/(4\zeta_1)+1/(4\zeta_2)-1]^2$ ($\approx
M_{B-L}^4[1/(4\zeta)-1]^2$), where $\zeta_1\equiv \rho_1
M_{B-L}^2/(\kappa_1 M_*^2)$, and $\zeta_2\equiv \rho_2
M_{B-L}^2/(\kappa_2 M_*^2)$. Eqs.~(\ref{sugravev})--(\ref{vac})
are valid only when $M_{B-L}^2|1/(4\zeta)-1|/2\lapproxeq |\langle
S\rangle|^2<<M_P^2$.\footnote{${\bf 16}_H$ and ${\bf
\overline{16}}_H$ develop the VEVs in the neutrino directions
$\langle\nu_H^c\rangle$, $\langle\bar{\nu}_H^c\rangle$. Near the
VEVs during inflation, the normalized real scalar fields,
Re$(\delta\nu_H^c+\delta\bar{\nu}_H^c)$ and
Im($\delta\nu_H^c-\delta\bar{\nu}_H^c$) acquire mass squareds
given by $m_{\pm}^2\approx 4\kappa^2|\langle S\rangle|^2\pm
2\kappa^2M_{B-L}^2[1/(4\zeta)-1]$, respectively~\cite{422}. } In
the limit $M_P\rightarrow \infty$, the above results approach the
values in global SUSY~\cite{422}.

Since $P$ and $Q$ develop VEVs, $X^{(')}$, $Y$, ${\bf 45}_H$, and
${\bf 45}_H'$ should also achieve VEVs from $D_{{\bf
16}^{(')}}W=D_{{\bf \overline{16}}^{(')}}W=0$ even during
inflation. Consequently, $SO(10)$ and $U(1)_A$ are broken to
$G_{SM}$ during inflation.
Note that
$\langle P\rangle=\langle \overline{P}\rangle$ and $\langle
Q\rangle=\langle \overline{Q}\rangle$ in Eqs.~(\ref{pvev}) and
(\ref{qvev}) lead to $\langle P_-\rangle=\langle Q_- \rangle=0$.
A non-zero vacuum energy from the ``F-term'' potential induces
universal ``Hubble induced scalar mass terms'' ($\kappa^2
M_{0}^4/M_P^2\times |\phi_l|^2$), which are read off from
Eq.~(\ref{inflpot}). But such small masses ($\kappa
M_{0}^2/M_P<<M_{B-L}$) cannot much affect the VEVs of the
superheavy scalars of order $M_{GUT}$.
%
%

Indeed, as seen earlier, in the SUSY limit the VEVs of $P$,
$\overline{P}$, $Q$, $\overline{Q}$ are not determined, even
though $\langle P\overline{P}\rangle$ and $\langle
Q\overline{Q}\rangle$ are fixed. But by including the SUSY
breaking soft terms of order $m_{3/2}$ in the scalar potential,
they are completely determined. Thus, one might expect that the
non-vanishing VEV of $S$ and the ``Hubble induced masses''
($>>m_{3/2}$) during inflation cause the VEVs of $P$,
$\overline{P}$, $Q$, $\overline{Q}$ to significantly deviate from
their values at low energies. Such differences, if true, would
result in oscillations by $P$, $\overline{P}$, $Q$, and
$\overline{Q}$ (or $P_{\pm}$ and $Q_{\pm}$) after inflation. As
explained earlier, with universal soft masses,
%
$\langle P_-\rangle=\langle Q_-\rangle=0$. Since the VEVs of $P_-$
and $Q_-$ vanish both during and after inflation, oscillations by
such light ($\sim m_{3/2}$) scalars would not arise after
inflation has ended.
%
%
%
%
%


A mass term for $S$ is induced by SUGRA corrections, such that the
``F-term potential'' contains
\begin{eqnarray}\label{sugracor}
V_F\supset \sum_{l} |D_{\phi_l}W|^2 \sim
\bigg(\frac{M_{GUT}}{M_P}\bigg)^2\times H^2|S|^2 ~,
\end{eqnarray}
where $H$ ($\equiv \kappa M_{\rm eff}^2/M_P \approx \kappa
M_{0}^2/M_P$) denotes the ``Hubble induced mass.''  Such a small
mass term of $S$ ($<<H^2|S|^2$) does not spoil the slow roll
conditions. The correction term in Eq.~(\ref{sugracor}) has a
small impact on the inflationary predictions.


With SUSY broken during inflation ($F_S\neq 0$), there are
radiative corrections from the ${\bf 16}_H$,
${\bf\overline{16}}_H$ supermultiplets, which provide logarithmic
corrections to the tree level potential
$V_F\approx\kappa^2M_{0}^4$, and thereby drive
inflation~\cite{hybrid}. In our model, the scalar spectral index
turns out to be $n_s=0.99\pm 0.01$ for $\kappa<10^{-2}$. (See FIG.
1.) The symmetry breaking scale $M_{B-L}$ is estimated to be
around $ 10^{16}-10^{17}$ GeV (FIG. 2).

Before concluding, some remarks about the reheat temperature
$T_r$, leptogenesis, and right handed neutrino masses are in
order.
When inflation is over, the inflatons decay into handed neutrinos.
Following Refs.~\cite{dec} and~\cite{LV},  a lower bound on $T_r$
is obtained for $\kappa =\lambda$, and the results are summarized
in FIG.3. (To obtain FIG.1-3, we set $M_*=M_P$ and
$\rho=\rho_{1,2}=1>>\frac{\kappa_{1,2}^2}{\kappa^2}$.) We see that
$T_r\lapproxeq 10^9$ GeV for $\kappa\lapproxeq 10^{-2}$.
The inflaton decay into right handed neutrinos yields the observed
baryon asymmetry via leptogenesis.
Assuming non-thermal leptogenesis and hierarchical right handed
neutrinos, we estimate the three right handed neutrinos masses to
be of order $10^{14}$ GeV, $(10-20)\times~T_r$ and few $\times~
T_r$. Note that with $\kappa < 10^{-2}$ the inflaton (with mass
$\sim\sqrt{\kappa M_{B-L}^2}$) can not decay into the heaviest
right handed neutrino (of mass $\sim 10^{14}$ GeV). Thus, the
latter does not play a direct role in leptogenesis.


In summary, our goal here was to realize inflation in a realistic
SUSY $SO(10)$ model. A global $U(1)_A$ and the $U(1)$ $R$-symmetry
play essential roles in the analysis. Several testable predictions
emerge. In particular, the scalar spectral index $n_s=0.99\pm
0.01$, which will be tested by several ongoing experiments. Proton
decay proceeds via $e^+ \pi^0$, with an estimated lifetime of
order $10^{34}-10^{36}$ yrs. The LSP is stable and the MSSM
parameter ${\rm tan}\beta$ is large, of order $m_t/m_b$. Two of
the three right handed neutrino masses are fairly well determined.
The heaviest one weighs around $10^{14}$ GeV, and the one
primarily responsible for non-thermal leptogenesis has mass of
order 10 $T_r$, where the reheat temperature $T_r$ is around
$10^8-10^9$ GeV.


\vskip 1.0 cm

\noindent {\bf Acknowledgments}

\noindent We thank Nefer Senoguz for helpful discussions and for
providing us with the figures. This work is partially supported
by the DOE under contract No. DE-FG02-91ER40626 (Q.S.).

\newpage

%
\begin{center}
\includegraphics[width=120mm]{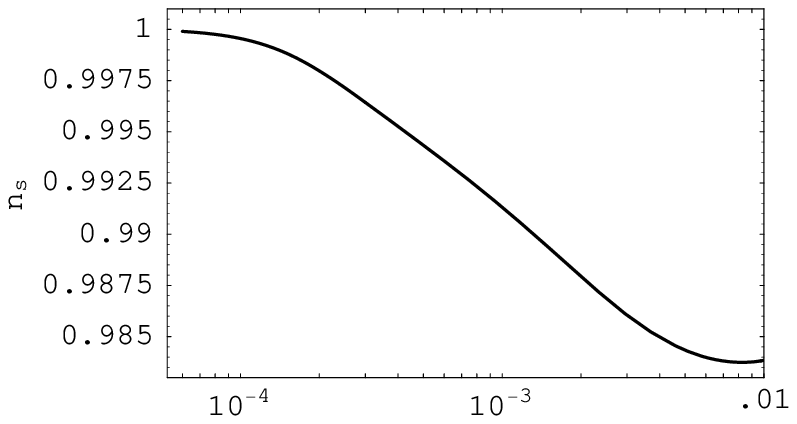}
\end{center}
FIG. 1: The spectral index $n_s$ vs $\kappa$. $\kappa$ is $<0.01$
so that the reheat temperature does $~~~~~~~$ not exceed $10^9$
GeV. See FIG. 3.  \vskip 2.5cm

\begin{center}
\includegraphics[width=120mm]{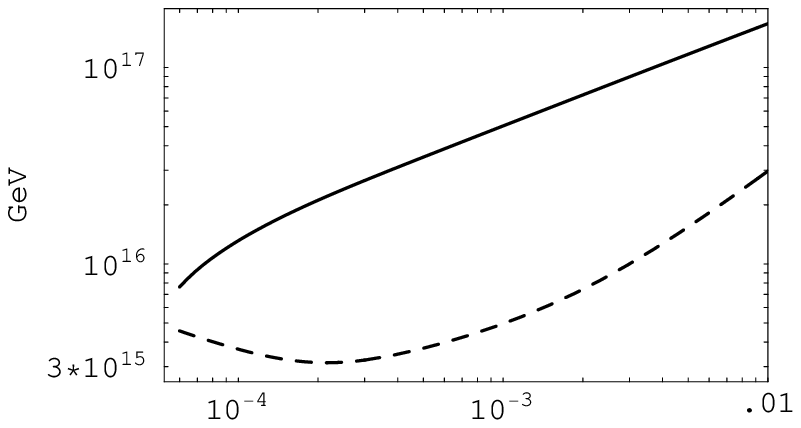}
\end{center}

FIG. 2: The symmetry breaking scale $M_{B-L}$ (solid) and
magnitude of the inflaton \\$~~~~~~~~~~~|S|$ (dashed) vs $\kappa$.

\begin{center}
\includegraphics[width=120mm]{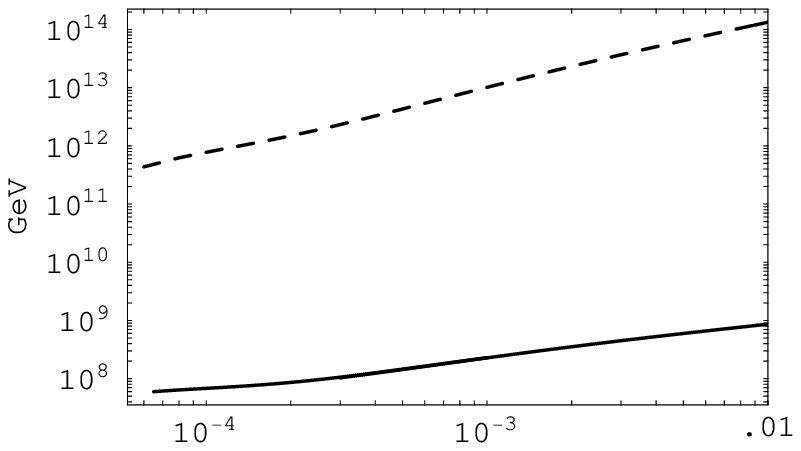}
\end{center}

~~~FIG. 3: Reheat temperature $T_r$ and inflaton mass (dashed) vs
$\kappa$.

\end{document}